\documentclass[aps,prl,twocolumn]{revtex4}
\usepackage{graphicx,amssymb,amsbsy,amsfonts,amssymb,amsmath}
\usepackage{hyperref}
\usepackage{multirow}
\usepackage{stackrel}
\usepackage{tikz}
\usetikzlibrary{calc}

\newcommand{\be}{\begin{equation}}
\newcommand{\ee}{\end{equation}}

\newcommand{\bea}{\begin{eqnarray}}
\newcommand{\eea}{\end{eqnarray}}

\newcommand{\eps}{\epsilon}
\def\eqn#1{eq.~(\ref{#1})}
\def\eqns#1#2{eqs.~(\ref{#1}) and (\ref{#2})}
\def\fig#1{fig.~{\ref{#1}}}

\def\nn{\nonumber}

\def\spa#1.#2{\left\langle#1\,#2\right\rangle}
\def\spb#1.#2{\left[#1\,#2\right]}
\def\spash#1.#2{\spa{\smash{#1}}.{\smash{#2}}}
\def\spbsh#1.#2{\spb{\smash{#1}}.{\smash{#2}}}
\def\sand#1.#2.#3{%
\left\langle\smash{#1}{\vphantom1}^{-}\right|{#2}%
\left|\smash{#3}{\vphantom1}^{-}\right\rangle}
\def\sandpp#1.#2.#3{%
\left\langle\smash{#1}{\vphantom1}^{+}\right|{#2}%
\left|\smash{#3}{\vphantom1}^{+}\right\rangle}
\def\sandpm#1.#2.#3{%
\left\langle\smash{#1}{\vphantom1}^{+}\right|{#2}%
\left|\smash{#3}{\vphantom1}^{-}\right\rangle}
\def\sandmp#1.#2.#3{%
\left\langle\smash{#1}{\vphantom1}^{-}\right|{#2}%
\left|\smash{#3}{\vphantom1}^{+}\right\rangle}

\def\ibp{IBP}

\def\r{\rho}

\def\BlackHat{{\sc BlackHat}}

\def\Mathematica{{\sc Mathematica}}
\def\Singular{{\sc Singular}}

\newbox\charbox
\newbox\slabox
\def\s#1{{      % Feynman slash
        \setbox\charbox=\hbox{$#1$}
        \setbox\slabox=\hbox{$/$}
        \dimen\charbox=\ht\slabox
        \advance\dimen\charbox by -\dp\slabox
        \advance\dimen\charbox by -\ht\charbox
        \advance\dimen\charbox by \dp\charbox
        \divide\dimen\charbox by 2
        \raise-\dimen\charbox\hbox to \wd\charbox{\hss/\hss}
        \llap{$#1$} }}

\begin{document}

\title{
\hbox{\rm\small
FR-PHENO-2017-007$\null\hskip 11.8cm \null$
UCLA-17-TEP-103
\break}
\hbox{$\null$\break}
    Two-Loop Four-Gluon Amplitudes with the Numerical Unitarity Method
}

\author{
S.~Abreu${}^a$, F.~Febres Cordero${}^a$, H.~Ita${}^a$, M.~Jaquier${}^a$, 
B.~Page${}^a$ and M.~Zeng${}^b$
\\
$\null$
\\
${}^a$ Physikalisches Institut, Albert-Ludwigs-Universit\"at Freiburg\\
       D--79104 Freiburg, Germany \\
${}^b$ Mani L.~Bhaumik Institute for Theoretical Physics \\
UCLA Department of Physics and Astronomy \\
Los Angeles, CA 90095, USA
}

\begin{abstract}
We present the first numerical computation of two-loop amplitudes based on the
unitarity method. As a proof of principle, we compute the four-gluon process. 
We discuss the new method, analyze its numerical properties and apply it
to reconstruct the analytic form of the amplitudes. The numerical
method is universal, and can be automated to provide
multi-scale two-loop computations 
for phenomenologically relevant signatures at hadron colliders.
\end{abstract}

\maketitle
The experiments at the Large Hadron Collider (LHC) at CERN are entering a new
phase in which observables will be studied with relative errors of the order of
a few percent. New discoveries through precision measurements require an
equal or better control over the theoretical uncertainties of predictions
from the Standard Model of particle physics.
A central bottleneck to obtaining such predictions is the complexity of
computing quantum corrections.
We demonstrate a new, automatable algorithm for computing two-loop corrections
in QCD based on the established unitarity method. In particular we
focus on the numerical variant of the method which has proven valuable for
dealing with multi-scale problems.
In addition to the flexibility of this approach, the geometric nature of
our method simplifies intermediate computational steps and promises good
numerical behavior.  
As a proof of principle of the new method we recompute the two-loop
gluon-gluon scattering amplitudes~\cite{4gluonGOT,4gluonBFD}.  This 
process exposes much of the complexity of two-loop
scattering amplitudes in QCD, including their universal infrared and
ultraviolet behavior.
The expressions in ref.~\cite{4gluonBFD} have been obtained using the analytic
variant of the unitarity method~\cite{Unitarity} which constructs scattering
amplitudes from their unitarity and analytic properties.
This method has been applied to a number of one-loop computations for the LHC,
and is the method of choice in formal research on scattering amplitudes in
supersymmetric theories.
In QCD, at two-loop level, the lack of (super)symmetry and the need for an
infrared and ultraviolet regulator pose further challenges due to the
appearance of one-loop sub divergences. 
Nevertheless, analytic computations of five- and six-gluon amplitudes
~\cite{AllPlusTwoLoop} have recently become available, albeit for constrained helicity
configurations.
Numerical variants of the unitarity approach at one-loop level
\cite{OPP,NumUnitarity,DNumUnitarity,BlackHat} have by now provided a large
number of phenomenologically relevant predictions for the LHC. The flexibility
of the numerical unitarity method in combination with its good numerical
stability motivate us to extend it to multi-loop amplitudes.
The algorithm which we put forward generalizes the one-loop approach
in a non-trivial way, as the two-loop
variant requires additional geometric input~\cite{IntDec}.
Being automatable and less susceptible to the complexity of analytic multi-scale
computations, our approach has the potential to mirror the successes found at
one-loop level.
In addition, analytic expressions can be efficiently reconstructed from a numerical
algorithm~\cite{Peraro:2016wsq}.

In this letter, we present the first numerical computation of two-loop QCD
amplitudes with the unitarity method. We focus on the
leading-color contributions to the two-loop four-gluon amplitudes which we
validate by comparing with known results~\cite{4gluonBFD}.
First, we set up the equations necessary for a hierarchical extraction
of an amplitude's
integrand~\cite{DProp}. Second, we decompose integrands of massless four-point amplitudes 
into master integrals and surface terms,  extending
the results of~\cite{IntDec}.
Third, we describe our numerical implementation and the linear algebra
techniques employed to compute coefficients for fixed values of the dimensional
regulator. 
Subsequently, we reconstruct the full regulator dependence and show
explicit numerical results.
Finally, we discuss numerical reconstruction of the analytic amplitude and give
our conclusions.\\

\begin{figure}[ht] \begin{tikzpicture}[scale=1.3]
% Level 1
\node at (4.6,3.3){\includegraphics[scale=0.13]{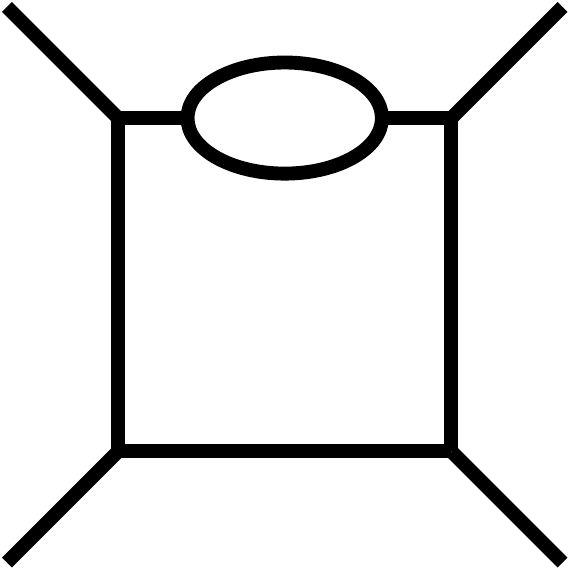}}; \node
at (2.8,3.3){\includegraphics[scale=0.13]{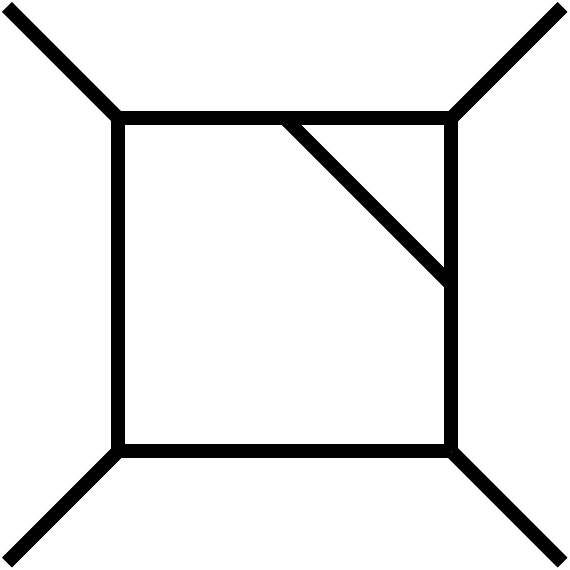}}; \node
at (1,3.3){\includegraphics[scale=0.13]{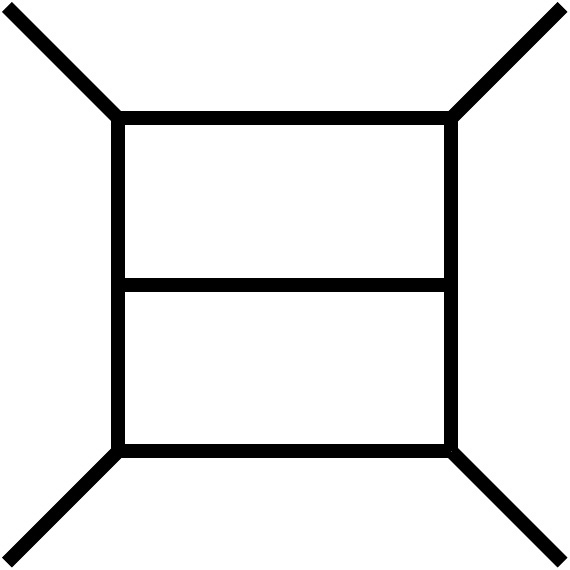}};
    % Level 2 
    \node at
(0.25,2.4){\includegraphics[scale=0.13]{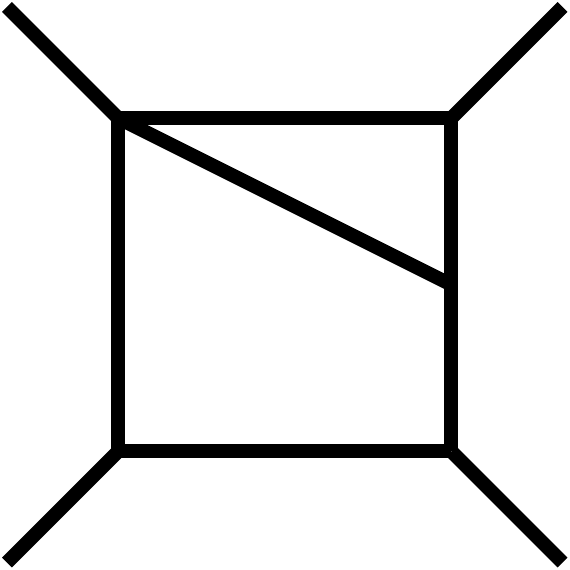}}; \node at
(1.1,2.4){\includegraphics[scale=0.13]{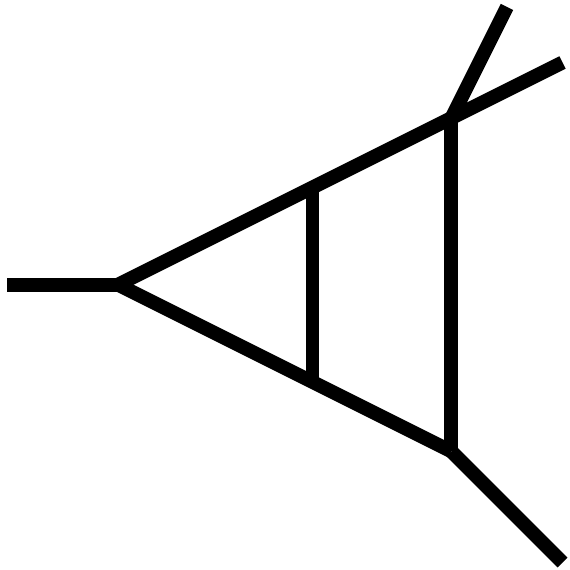}}; \node
at (1.95,2.4){\includegraphics[scale=0.13]{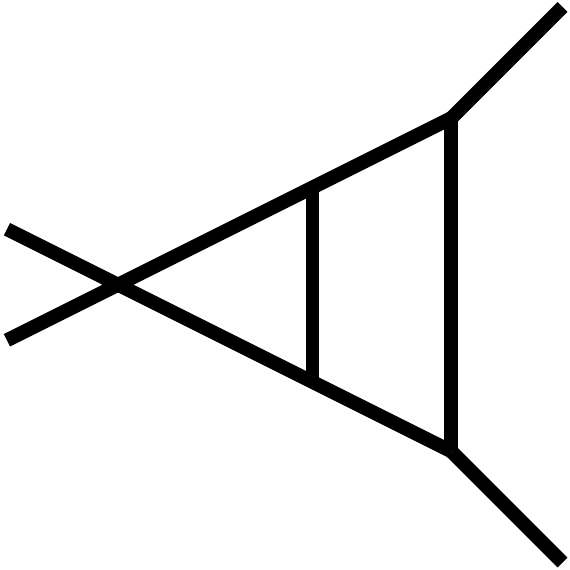}};
\node at
(2.8,2.4){\includegraphics[scale=0.13]{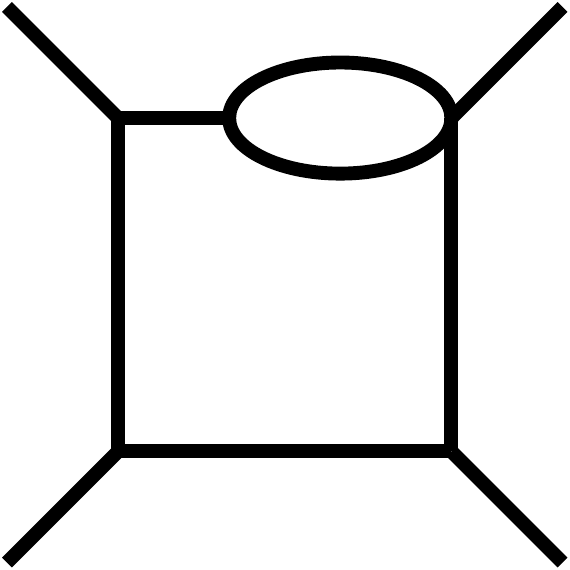}};
\node at
(3.65,2.4){\includegraphics[scale=0.13]{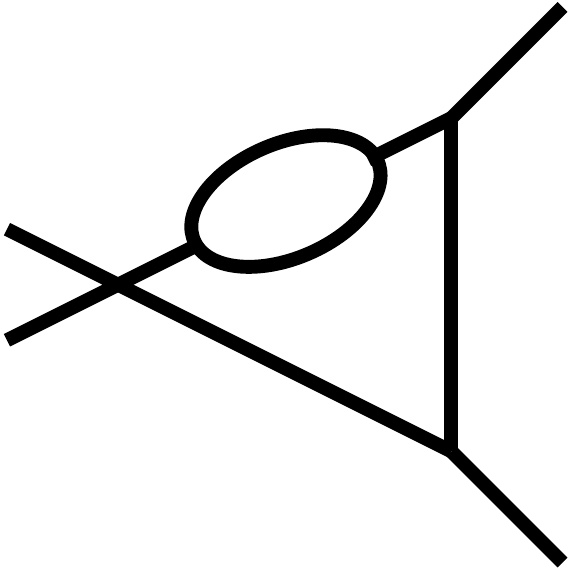}};
\node at
(4.5,2.4){\includegraphics[scale=0.13]{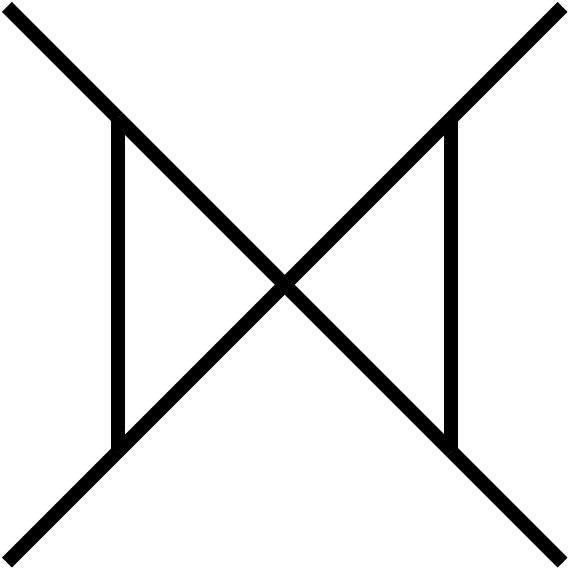}};
\node at
(5.35,2.4){\includegraphics[scale=0.13]{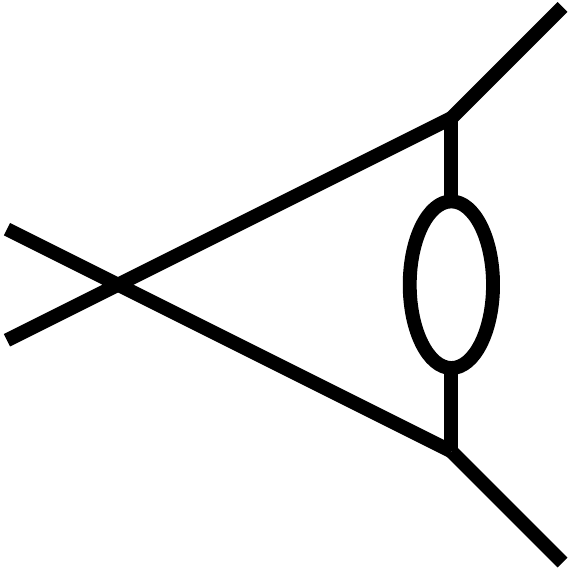}};
    % Level 3
    \node at
(0,1.4){\includegraphics[scale=0.13]{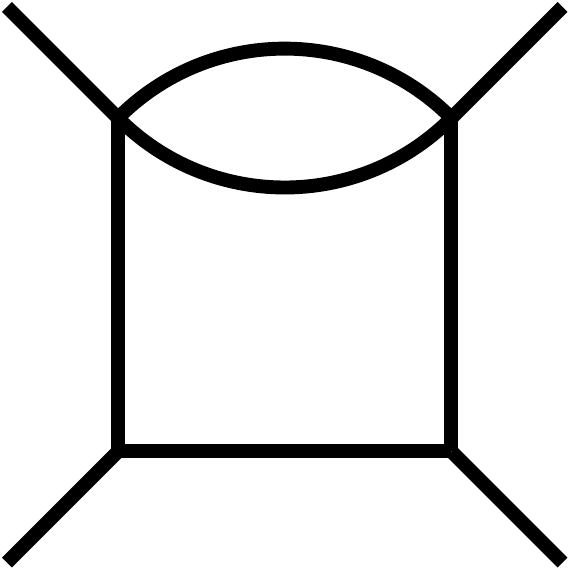}}; \node at
(.7,1.4){\includegraphics[scale=0.13]{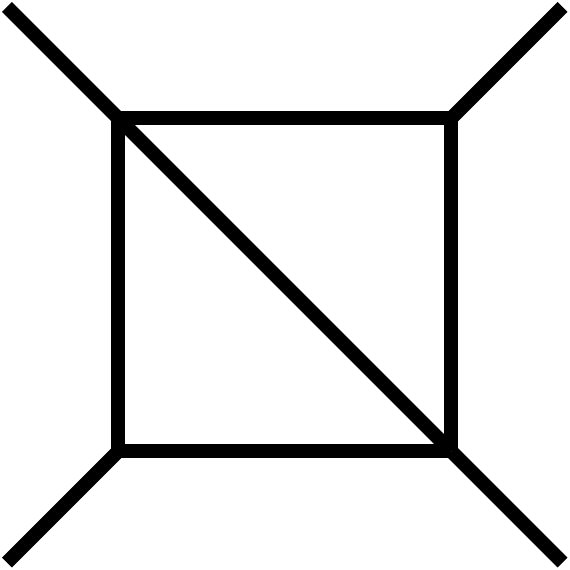}};
\node at
(1.4,1.4){\includegraphics[scale=0.13]{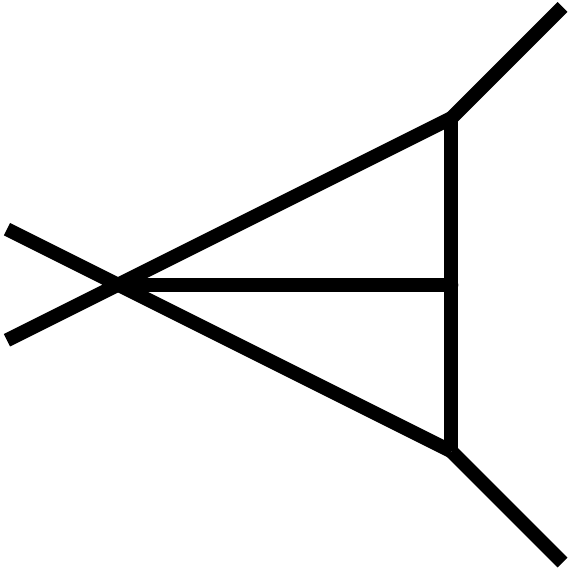}};
\node at
(2.1,1.4){\includegraphics[scale=0.13]{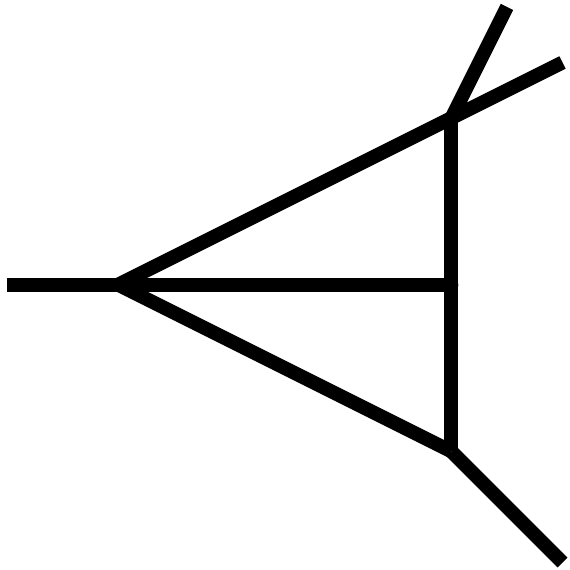}};
\node at
(2.8,1.4){\includegraphics[scale=0.13]{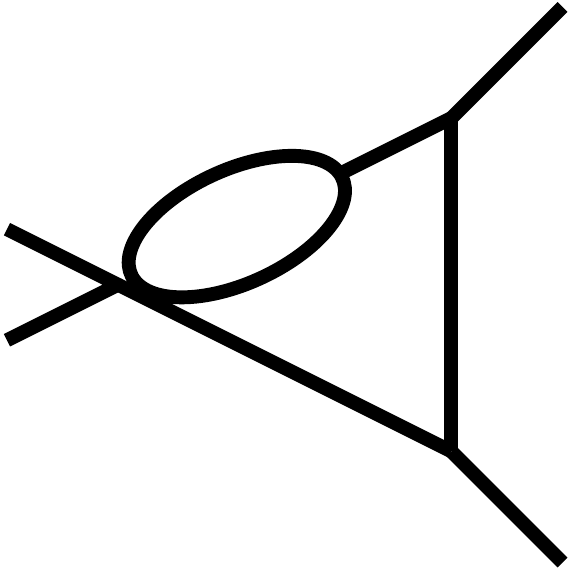}};
\node at (3.5,1.4){\includegraphics[scale=0.14,angle=180]{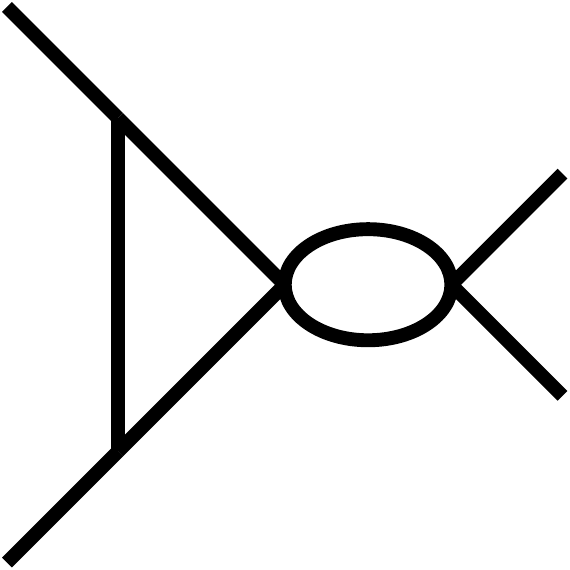}};
\node at (4.2,1.4){\includegraphics[scale=0.13]{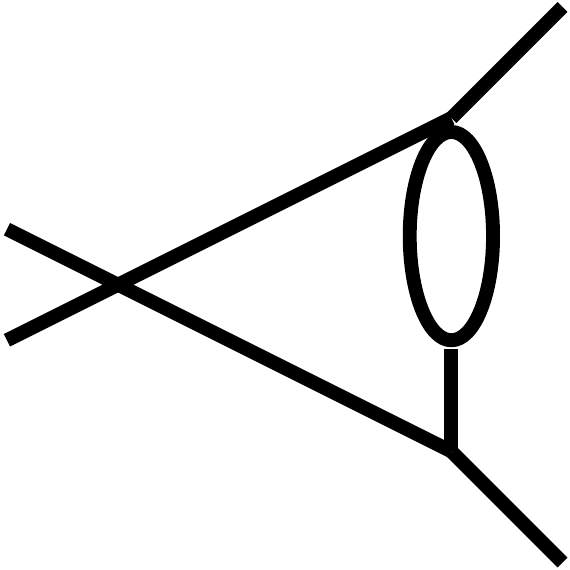}};
\node at (4.9,1.4){\includegraphics[scale=0.14]{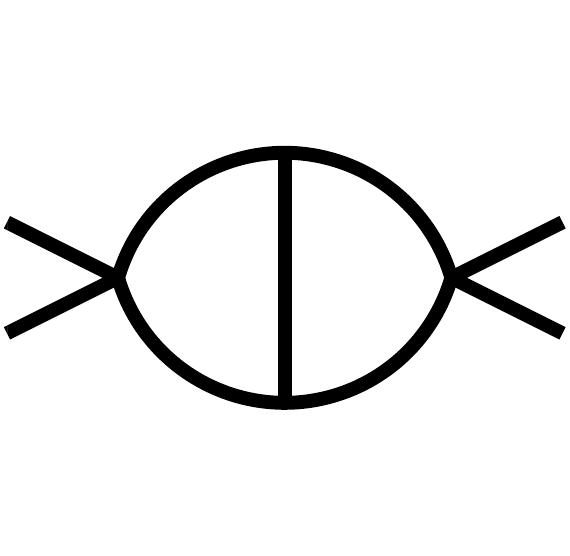}};
\node at (5.6,1.4){\includegraphics[scale=0.14]{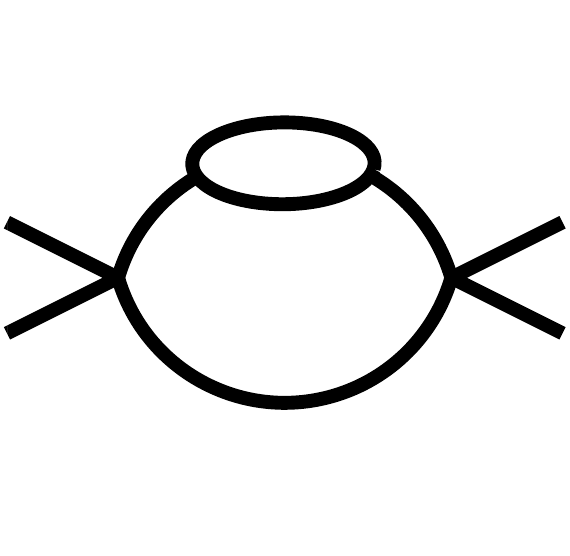}};
    % Level 4
    \node at 
(4.6,.6){\includegraphics[scale=0.13]{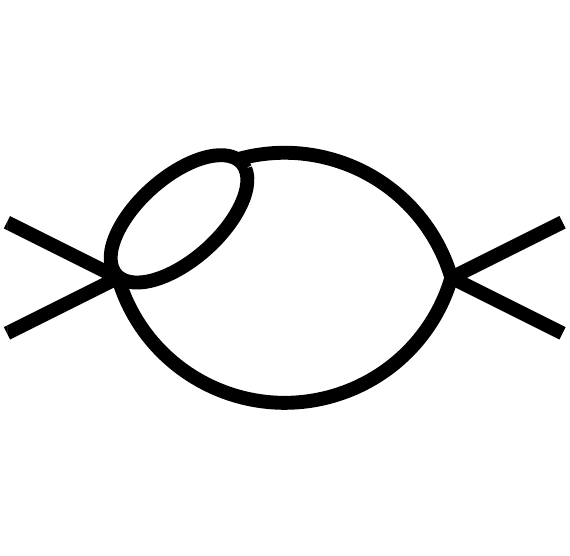}};
    \node at
(3.4,.6){\includegraphics[scale=0.13]{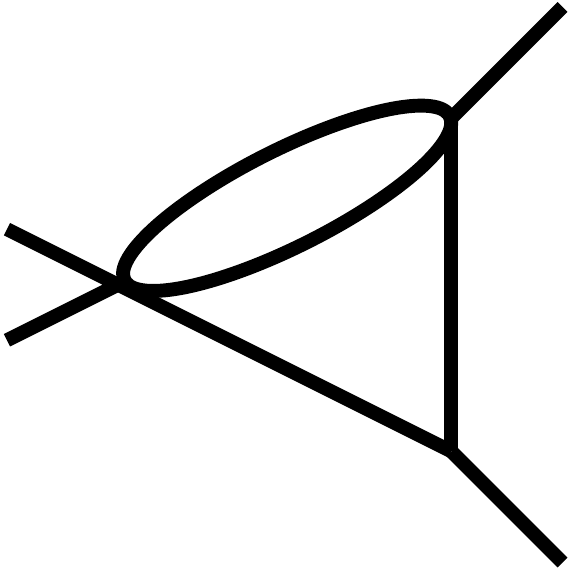}};
\node at 
(2.2,.6){\includegraphics[scale=0.13]{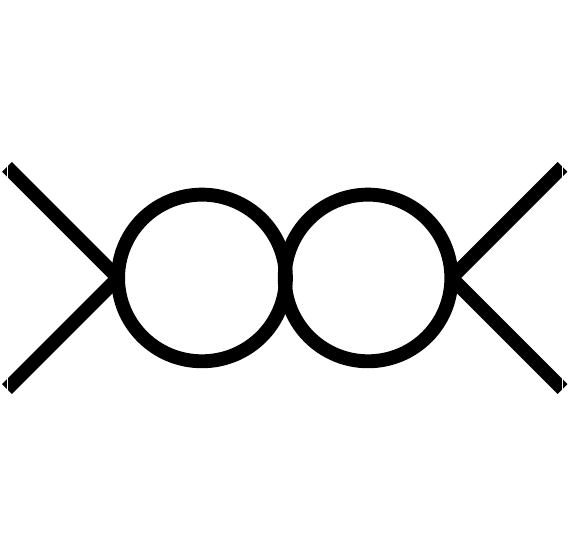}};
\node at
(1,.6){\includegraphics[scale=0.13]{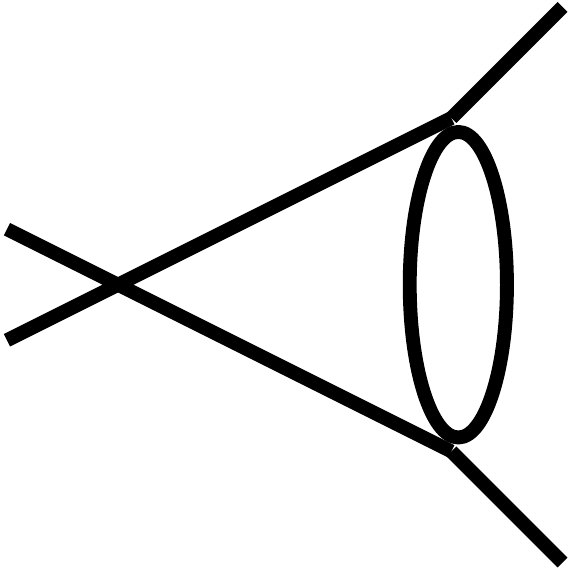}}; 
    % Level 5
    \node at (2.8,-0.1){\includegraphics[scale=0.15]{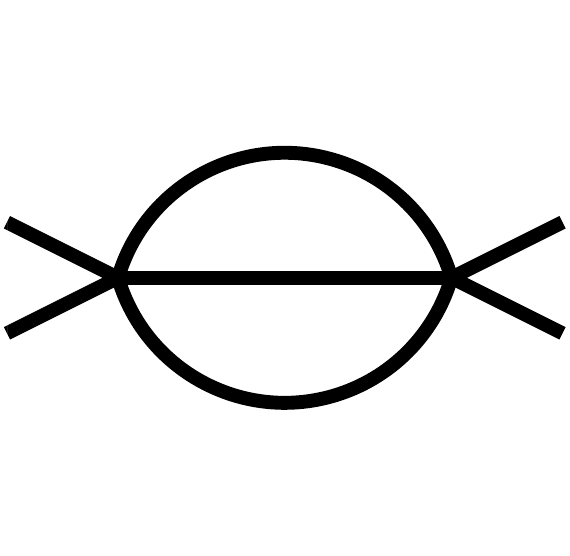}};
\end{tikzpicture} \caption{The hierarchy of propagator structures for
planar four-point two-loop gluon amplitude. Only topologically inequivalent
structures are shown.} 
\label{diagSunrise} 
\end{figure}

\noindent{\bf Numerical Unitarity Method.}
We apply a variant of the unitarity method suitable for analytic and numerical computations
which generalizes one-loop methods
to higher loop orders. 
For more details of our approach we refer the reader to ref.~\cite{DProp}.
We start with an ansatz for the integrand ${\cal
A}(\ell_l)$ of a two-loop amplitude~\cite{IntDec},
\begin{equation}
{\cal A}(\ell_l)=\,\sum_{\Gamma \in \Delta}\,\,\sum_{i\,\in\, M_\Gamma\cup S_\Gamma}
\frac{ c_{\Gamma,i} \,m_{\Gamma,i}(\ell_l)}{\prod_{j\in P_\Gamma} \rho_j}\,,
\label{eq:AL} 
\end{equation}
where $\Delta$ denotes the set of two-loop diagrams
 which specify the possible propagator structures of the
amplitude.
In \fig{diagSunrise} we show the set $\Delta$ corresponding to
planar two-loop four-point massless amplitudes.
The set of diagrams $\Delta$ is organized hierarchically with the partial
ordering
$\Gamma_1>\Gamma_2$ if $P_{\,\Gamma_1}\supset P_{\,\Gamma_2}$,
i.e., if the diagram $\Gamma_2$ (a descendant) is obtained
from $\Gamma_1$ (an ancestor) by
removing one or more propagators.
Here $P_\Gamma$ denotes the set of propagators
associated to diagram $\Gamma$. 
The loop momenta 
are denoted by $\ell_l,\,l=1,2$ and inverse propagators by~$\rho_j$.
The numerators span the full set of independent
numerator terms~\cite{NumPar} and are restricted by power counting. 
For each diagram, the set $M_\Gamma$ specifies numerators associated to
master integrals and $S_\Gamma$ the set of surface terms
(i.e., terms that integrate to zero) necessary to fully
parameterize the numerators.
The $m_{\Gamma,i}(\ell_l)$ are polynomials in~$\ell_l$. We work in dimensional
regularization such that these, as well as the coefficient functions
$c_{\Gamma,i}$, depend on the loop-momentum dimension
$D=4-2\epsilon$. 
Dependence on external kinematics is implicit.
For simplicity we omit color information and focus on leading-color
contributions.

%%%%%%%%%%%%% FIGURE %%%%%%%%%%%%%%%%%%
%%%%%%%%%%%%%%%%%%%%%%%%%%%%

The physical scattering
amplitude is obtained in terms of master integrals~$I_{\Gamma,i}$,
\begin{eqnarray}
{\cal A}&=&\sum_{\Gamma\in \Delta} \sum_{i\,\in\, M_\Gamma} c_{\Gamma,i}
I_{\Gamma,i}\,, \label{eq:A} 
\end{eqnarray}
as surface terms drop out from \eqn{eq:AL} after integration.
At this stage, we only need to compute the coefficient functions $c_{\Gamma,i}$.
In generalized unitarity, they
are determined by solving the linear system of equations \eqref{eq:AL} for values
of the loop momenta $\ell_l^\Gamma$ for which the internal particles go
on-shell, $\rho_j \rightarrow 0$ for all $j\in P_\Gamma$. %
%
%%%%%%%%%%%%%%%%%%%%%%%%%%%%%%%%%%
%
In the absence of subleading poles associated with this limit~\cite{DProp},
putting the ansatz (\ref{eq:AL}) and factorization limits %(\ref{eq:UE}) 
together we obtain
\begin{eqnarray}
\sum_{\rm states}\prod_{i\in T_\Gamma} {\cal A}^{\rm tree}_i(\ell_l^\Gamma) =
\sum_{\substack{\Gamma' \ge \Gamma\,,\\ i\,\in\,M_{\Gamma'}\cup S_{\Gamma'}}
} \frac{ c_{\Gamma',i}\,m_{\Gamma',i}(\ell_l^\Gamma)}{\prod_{j\in
(P_{\Gamma'}\setminus P_\Gamma) } \rho_j(\ell_l^\Gamma)}\,, \label{eq:CE} 
\end{eqnarray}
which we call the {\it cut equations}. The set $T_\Gamma$ labels all 
tree amplitudes corresponding to the vertices in the diagram~$\Gamma$.
The state sum runs over all possible internal states, here the $(D_s-2)$
gluon helicity states for each internal
line of $\Gamma$. 

The cut equations apply only for the subset of diagrams
$\Gamma\in \Delta'$ for which the factorization limit has no subleading
poles, i.e., not all propagator structures yield a cut
equation. 
Nevertheless, it is possible to organize the set of cut equations to produce enough linear
systems of equations to compute all coefficient functions $c_{\Gamma,i}$~\cite{DProp}.

This variant of the unitarity approach for computing multi-loop amplitudes is
suitable both for analytic and numerical calculations.
In what follows, we will focus on its numerical implementation.\\

\noindent {\bf Construction of integrand parameterization.}
For the construction of the integrand (\ref{eq:AL}) we extend the method of
ref.~\cite{IntDec} to massless propagators and external momenta.
The ansatz (\ref{eq:AL}) relies on a suitable
set of integration-by-parts (\ibp) relations,
\begin{eqnarray}
0 = \int \prod_{l=1,2} d^D\ell_l \frac{\partial}{\partial \ell_{j}^\nu} \left[
    \frac{ u_{j}^\nu }{\prod_{k\in P_\Gamma} \rho_k}\right],  \label{eq:ibp}
\end{eqnarray}
which we use to construct the numerators $m_{\Gamma,i}(\ell_l)$. 
In order to control propagator powers to match the ones of the integrand
(\ref{eq:AL}), we construct \ibp{} relations from constrained vector
fields~\cite{IBPGKK} which solve,
\begin{eqnarray}
\label{eq:GKK} u_i^\nu \frac{\partial}{\partial \ell_{i}^\nu} \rho_{j}=
f_j \rho_j, 
\end{eqnarray}
where no summation over the index $j$ is implied. Solutions $\{u_i^\nu\}$ of
\eqn{eq:GKK} are referred to as \ibp{}-generating vectors.
Alternatively, reduction programs \cite{IntDecProg} can be used to obtain
appropriate \ibp{} relations.

%%%%%%%%%%%%% FIGURE %%%%%%%%%%%%%%%%%%
\begin{figure}[ht] 
   \includegraphics[scale=0.55]{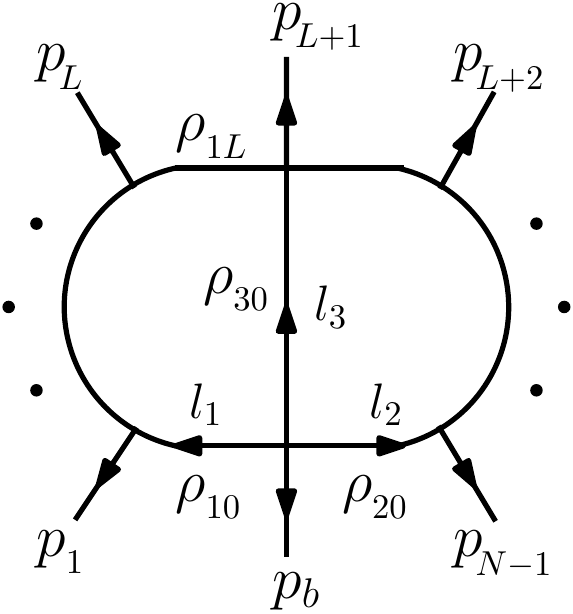}
   \caption{ Displayed are the conventions for assigning propagators in a
   two-loop diagrams.} 
\label{fig:diagConventions} 
\end{figure}
%%%%%%%%%%%%%%%%%%%%%%%%%%%%

An important tool for solving \eqn{eq:GKK} and analyzing
\ibp{} relations are the natural coordinates \cite{Baikov,IntDec,LZ},
\begin{eqnarray}
&&\!\!\!\!\!\!\!\!\!\!\!\!\!\!\!\!
\ell_l =\sum_{j\in B_l^p} v_l^j r^{lj} + \sum_{j\in B_l^t} v_l^j \alpha^{lj} +
\sum_{i\in B^{ct}} n^i \alpha^{li} + \sum_{i\in B^\epsilon} n^i \mu_l^i
\,,\label{eq:moms}\\ 
&&\!\!\!\!\!\!\!\!\!\!\!\!\!\!\!\!
r^{li}= -\frac{1}{2}(\rho_{li} - (q_{li})^2 - \rho_{l(i-1)} + (q_{l(i-1)})^2)\,,
\end{eqnarray}
which parameterize the strand momenta $\ell_l$ ($l=1,2,3$, see
\fig{fig:diagConventions}) in terms of inverse propagators
$\rho_{li}$, and auxiliary variables $\alpha^{li}$ and $\mu_l^i$.
The vectors $q_{li}$ are fixed by momentum conservation after imposing $q_{10}=q_{20}=0$.
The vectors $n^i$ form an orthonormal basis transverse to the scattering plane,
i.e~$n^i\cdot p_j=0$.  Labels in $B^\epsilon$ refer to directions beyond four
dimensions and labels in $B^{ct}$ denote transverse directions within four
dimensions.
Within the scattering plane, we setup distinct basis vectors $v_l^i$ for each
strand of the diagram; the vectors $p_i$ with ${i\in B_l^p}$ which
exit the strand $l$ are completed with additional physical momenta $p_i$ with
${i\in B_l^t}$, so as to span the whole scattering plane.
The vectors $v_l^i$ are dual to the $p_j$, $v_l^i= (G_l)^{ij}
p_j$, with $(G_l)^{ij}$ the inverse of the Gram matrix $(G_l)_{ij}=p_i\cdot
p_j$. 
The new coordinates $\alpha^{li}$, $\mu_l^i$ and $\rho_{li}$ are not independent:
momentum conservation $\ell_1+\ell_2+\ell_3 +p_b=0$ removes redundant
$\alpha_{li}$ and $\mu_3^i$ variables. We often label $\r$ variables and
independent $\alpha$ variables by a single subscript referring to the pair of
strand index $l$ and label $i$.
Furthermore, we have the constraint
\begin{eqnarray} \label{eq:mus}
\mu_{ll}\equiv(\mu_l)^2 &=& \rho_{l0}- \sum_{\nu=0}^3 \ell_l^\nu \ell_{l\nu}\,
\end{eqnarray}
for the extra-dimensional momentum squares.

In these coordinates,
\ibp-generating vectors are \cite{IntDec}
\begin{eqnarray} \label{eq:ugen}
u= f_{i}\r_{i}\frac{\partial}{\partial\r_{i}} + u_j \frac{\partial}{\partial
\alpha^j}+\sum_{l,l'=1,2}f^l_{l'} \mu^k_l \frac{\partial}{\partial
\mu^k_{l'}}\,,
\end{eqnarray}
where we denote basis vectors along coordinate lines by the respective partial
derivatives. Repeated indices are summed over. We impose rotation symmetry in
the extra-dimensional components as manifest in the integrand (\ref{eq:AL}).
The vectors must be polynomial when written in canonical coordinates of $\ell_l$,
requiring first that the coordinate functions $f_i$, $f_l^{l'}$ and
$u_i$ to be polynomial in $\alpha^i$ and $\rho_i$, and second that the vectors are
compatible with the relations in \eqn{eq:mus}. 
After redundant variables have been eliminated, the compatibility conditions
are quadratic in $\alpha^i$ and $\rho_i$, and read
\begin{align}\begin{split}\label{eq:syz1}
\bar u(\mu_{11} ) &= 2 \mu_{11} f_1^1 + 2 \mu_{12} f_1^2 \,, \\
\bar u(\mu_{22} ) &=  2 \mu_{22} f_2^2 + 2 \mu_{12} f_2^1 \,, \\
\bar u(\mu_{12}) &= \mu_{12} (f_1^1 + f_2^2) + \mu_{11} f_2^1 + \mu_{22}
f_1^2\,.
\end{split}\end{align}
We use $\bar u$ to denote the $\mu$-independent
directions of $u$ in eq.~(\ref{eq:ugen}) and $\mu_{12}\equiv(\mu_{33}-\mu_{11}-\mu_{22})/2$.
The generating set of solutions for the unknowns $f_i, f_{l'}^l$ and $u_j$ are
obtained using \Singular{} \cite{Singular} which we use in addition to
universal vectors \cite{IntDec}.  We consider the solutions that do not vanish
on-shell and keep a generating set of these. An automated implementation of
the on-shell variant of the above equations is also
available~\cite{Azurite}.  

The simplest IBP-relations are obtained from diagonal rotations \cite{IntDec}.
The respective relations are generalizations of the one-loop numerators
\cite{OPP,NumUnitarity} being traceless completions of the transverse
monomials. They can be systematically constructed from traceless
completions
\begin{align} 
\begin{split}
&    \alpha^{lj}\alpha^{l'j}-b_1\frac{\mu_{ll'}}{\epsilon}\,,\quad
    l,l'\in\{1,2\},\\
&    
    (\alpha^{lj})^2(\alpha^{l'j})^2\!-\!
    b_1\frac{\mu_{ll}}{\epsilon}(\alpha^{l'j})^2\!-\!b_2\frac{\mu_{ll'}}{\epsilon}\alpha^{lj}\alpha^{l'j}\,,\,
    l\ne l'\,,
\end{split}\label{eq:TC}
\end{align}
with $j\in B^{ct}$. After multiplying with additional $\alpha$'s either from
$B^t_l$, $B^t_{l'}$ or $B^{ct}$ the rational constants $b_1$ and $b_2$ are
fixed to yield expressions that are transverse to the Lorentz structures after
loop integration. Monomials with odd powers of $\alpha$'s from $B^{ct}$
integrate to zero and can be used for numerators as they are.
The completion of the \ibp{} relations is obtained by multiplying the
\ibp-generating vectors with the independent $\alpha$ variables.  The
numerators are,
\begin{eqnarray} m_{\Gamma,u}(\ell_l) &=& \left[-(\nu_i-1) f_i + \rho_{i}
        \frac{\partial f_{i}}{\partial \rho_{i} } + \frac{\partial
            u_{j}}{\partial \alpha^j} +\right.\nn\\ &&\left.\left(D -
            \frac{n_\alpha+1}{2}\right)(f_1^1+f_2^2) \right]\,, \label{eq:m} \end{eqnarray}
taking into account the Jacobian factor from \eqn{eq:ibp}, see
e.g.~\cite{Zhang:2016kfo}.  In \eqn{eq:m}, $n_\alpha$ denotes the number of
independent $\alpha_i$-variables and $\nu_i$ the propagator power.  

The numerator decomposition into surface terms and master integrands
must be non redundant and complete.  The spanning set of \ibp-relations is
obtained modulo inverse propagators, which allows to check the linear
independence with on-shell conditions $\rho_i=0$ imposed.
Finally, the numerators of master integrands are given by the span of
irreducible numerators modulo surface terms. In natural
coordinates, the irreducible numerators are the monomials in $\alpha^i$ 
consistent with power counting. It is sufficient to compare the span of
\ibp-relations and irreducible numerators on a given phase-space point with
on-shell conditions imposed to determine a decomposition in terms of master integrals
as in \eqn{eq:A}~\cite{IntDec,LZ}.\\

\noindent{\bf Numerical implementation.}
We first decompose the amplitude into master integrals
and surface terms as in \eqn{eq:AL} with the method described above.
We find two master integrals associated
with the first diagram in the first line of \fig{diagSunrise} (the {\it double-box})
and one master integral associated with each of the first two diagrams of the third line,
the first two diagrams of the fourth line, and the diagram in the last line.
For the double-box diagram we choose the scalar and the (irreducible) numerator
insertion, and for all other diagrams we choose the scalar integrals.

The coefficients $c_{\Gamma,i}$ in eq.~(\ref{eq:CE}) must now be fixed.
They are determined numerically
by sampling of the on-shell phase spaces $\ell_l^\Gamma$,
for all $\Gamma\in\Delta'$~\cite{DProp}.
This was implemented in a
C++ library, for which the required analytic
information - color decomposition (see ref.~\cite{Color2Loop}), hierarchy of
cut equations organized to handle subleading poles and
\ibp{} relations - is produced in \Mathematica{}. For the four-dimensional
spinor algebra we have used tools from the \BlackHat{} library~\cite{BlackHat}.

We built an (over)constraining system of linear equations for
$c_{\Gamma,i}$ by computing the subtracted product of trees (through off-shell
recursions~\cite{BGrec}) at randomly sampled phase-space points. Though
analytically $D$-dimensional, phase-space samples can be constructed with
6-dimensional momenta due to the rotational invariance beyond the 4-dimensional
physical slice.  The on-shell phase spaces are generated by nested one-loop
parameterizations~\cite{BlackHat,NumUnitarity}.  We single out one of the
loops and construct loop momenta on its on-shell phase space.
We then input these momenta into the second
loop to find two-loop on-shell configurations.  

In order to numerically solve the linear systems we constructed for the
$c_{\Gamma,i}$, we employ standard linear
algebra techniques as implemented in the \texttt{LAPACK} library~\cite{LAPACK}
which uses the \texttt{BLAS} routines~\cite{BLAS}. For
higher-precision arithmetics, we also use the associated routines from the
\texttt{MPACK} libraries~\cite{MPACK}.
For every $n\times n$ system of
equations $Mc^{}=a^{}$
we employ a $PLU$
factorization of the square matrix $M$, in which $L$ is a lower-triangular
matrix, $U$ is an upper-triangular matrix and $P$ is a permutation matrix.
For over constrained systems of equations, we employ a $QR$
factorization ($Q$ a rectangular orthogonal matrix, and $R$ a squared upper-triangular one)
for minimizing $||Mc^{}-a^{}||$. These factorizations
allow for
efficient and numerically stable solutions to the systems of
equations, which are typically order 100-dimensional. 

By solving these systems, we numerically determine the coefficients of master
integrals for fixed values of $D$ and $D_s$. The $D_s$ dependence of
master-integral coefficients is at most quadratic
and can be reconstructed \cite{DNumUnitarity} by evaluating the
coefficients with state sums (\ref{eq:CE}) for three different
values of $D_s$. The $D$ dependence of the coefficients  is rational and
originates from the $D$-dependent \ibp{} relations, see~\eqns{eq:TC}{eq:m}.
A priori, the polynomial degrees of the numerator and denominator
of the rational function are unknown.
However, as this is phase-space independent, it can be determined once and for all
in a dedicated run using rational
reconstruction techniques  in high precision \cite{Peraro:2016wsq,Abramowitz}.
Further, because the coefficients of the denominators are rational
numbers, they can be exactly reconstructed using continued
fractions. In this way, after a warm-up phase, one needs only reconstruct a
polynomial dependence of known rank.

Finally, to obtain values for the amplitudes
(\ref{eq:A}) we
combine the integral coefficients with our own implementation of the
loop integrals~\cite{Integrals4pt} using the function library of
ref.~\cite{HPL}. By setting $D_s = D = 4-2\epsilon$ in the
coefficients and expanding around $\eps=0$, we recover the value of the bare amplitude
as a Laurent series in $\epsilon$ in the HV~\cite{HV} variant of 
dimensional regularization.\\

\noindent {\bf Validation of results.}
We have computed the two
independent helicity configurations with non-vanishing tree amplitudes
in the leading-$N_c$ color limit,
omitting  contributions from closed fermion loops.
We normalize the results to the respective tree-level amplitudes ${\cal A}_0$.
We provide numerical values for
the coefficients of the Laurent series in $\eps$ in the HV scheme. 
For the point $p_{1,3}=-1/2\,(1,0,0,\pm 1)$,
$p_{2,4}=1/2\,(1,0,\pm \sqrt{3}/2,\pm 1/2)$, we find the following values for the bare 
amplitudes ($g_s=1$ and $\mu=1$):
{\small
\begin{center}
\begin{tabular}{|c||c|c|c|c|c|} \hline
    ${\cal A}/({\cal A}_{0}N_c^2 )(4\pi)^4 $ 
    & $\epsilon^{-4}$ & $\epsilon^{-3}$ & $\epsilon^{-2}$ & $\epsilon^{-1}$ & $\epsilon^0$ \\ \hline\hline 
    $(1_g^-,2_g^+,3_g^-,4_g^+)$ & 8.00000 & 55.6527 & 176.009 & 332.296 & 486.502 \\[.5mm]\hline 
    $(1_g^-,2_g^-,3_g^+,4_g^+)$ & 8.00000 & 55.6527 & 164.642 & 222.327 & -8.39044 \\[.5mm]\hline 
\end{tabular}
\end{center}
}
\noindent We have checked that the above results agree with the expected universal infrared-pole structure
\cite{CataniFormula}, and that they match the results obtained
from the analytic expressions of ref.~\cite{4gluonBFD}.
These comparisons validate the C++ implementation of our numerical
unitarity algorithm.

\begin{figure}[h]
\includegraphics[scale=0.7]{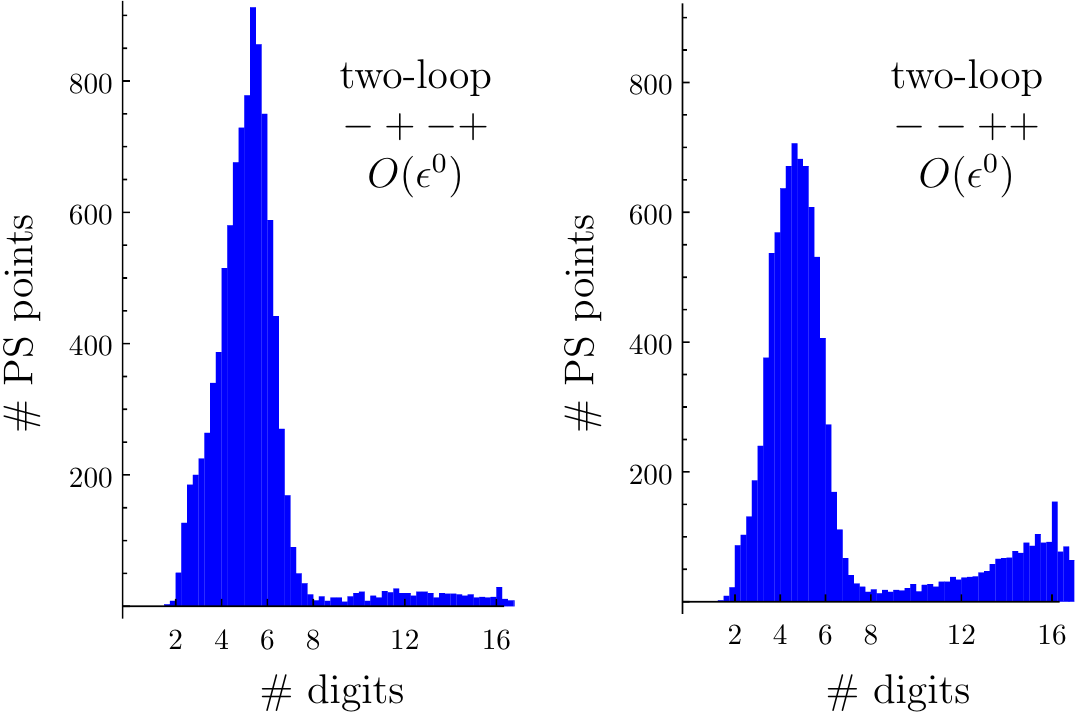}
\caption{Distribution of minus the base 10 logarithm of the relative error for
the numerical calculation with respect to the analytic result, over 10,000
phase-space points. The distribution corresponds to the finite ${\cal
O}(\epsilon^0)$ contribution. The left plot is for the $(1_g^-,2_g^+,3_g^-,4_g^+)$
helicity configuration, and the right plot for $(1_g^-,2_g^-,3_g^+,4_g^+)$.
}
\label{fig:stability}
\end{figure}

In \fig{fig:stability}, we show the stability of our calculation by looking at
minus the base 10 logarithm of the relative error for the numerical calculation
with respect to the analytic result. This is done over 10,000 phase-space points
evenly distributed, with a minimum $p_T$ cut on the final-state partons set to $1/100$ the total energy in the center-of-mass frame.
The numerical computation has employed single-double precision
operations, except when extracting the maximal numerators for which we choose to
use double-double precision arithmetics. We have also introduced a rescue
system, based on comparisons to the known pole structure of the
amplitude~\cite{CataniFormula}, and
when failing we recompute the full amplitude in higher floating-point precision. We
observe that our calculation is precise to better than a per mill relative error
for the bulk of phase space and that the rescue system works properly by shifting 
the left tail of the distribution towards 16 digits to the right.
When computing with single-double precision over phase space and after
reconstruction of the dimensional regulators, our results for master integral
coefficients are commonly accurate to 8 digits.

Finally, we have produced results with quadruple-double precision~\cite{QD},
which gives enough numerical precision to fully reconstruct the analytic
form of the amplitudes from the purely numerical computations. This is achieved with the
same techniques as for the regulator reconstruction since, after
rescaling, the integral coefficients are a function of a single variable.\\

\noindent {\bf Conclusions.}
We have presented a method based on generalized unitarity for automated analytic
and numerical computations of two-loop scattering amplitudes in QCD. We have
validated our algorithm by comparing the numerical results to the ones of known
analytic amplitudes. The method is numerically stable and refinements can
improve this further in the future. Furthermore, our method can be
used to extract the analytic expressions for the amplitude from the purely
numerical results. 
In the last decade, similar developments at one-loop level have led to important
predictions for high-multiplicity processes at hadron colliders. We are thus
optimistic that our approach will be instrumental in yielding new two-loop matrix
elements necessary for QCD phenomenology in the near future.\\

\noindent {\bf Acknowledgements:} We thank Z.~Bern, L.D.~Dixon and D.A.~Kosower for
many helpful discussions. We particularly thank Z.~Bern for providing analytic
expressions from ref.~\cite{4gluonBFD}.
We thank C.~Duhr for the use of his \Mathematica{} package \texttt{PolyLogTools}.
We thank J.~Dormans and E.~Pascual
for technical discussions.
H.I is grateful to the Mainz Institute for Theoretical Physics (MITP)
for its hospitality and its partial support during the completion of this work.
S.A.'s work is supported by the Juniorprofessor Program of Ministry of Science, 
Research and the Arts of the state of Baden-W\"urttemberg, Germany.
H.I.'s work is supported by a Marie Sk{\l}odowska-Curie Action
Career-Integration Grant PCIG12-GA-2012-334228 of the European Union.  The work
of F.F.C., M.J. and B.P. is supported by the Alexander von Humboldt Foundation, in the
framework of the Sofja Kovalevskaja Award 2014, endowed by the German Federal
Ministry of Education and Research. 
The work of M.Z. is supported by the U.S. Department of Energy under Award Number
DE-{S}C0009937.
This work was performed on the
bwUniCluster funded by the Ministry of Science, Research and the Arts
Baden-W\"urttemberg and the Universities of the State of Baden-W\"urttemberg,
Germany, within the framework program bwHP.

\end{document}